\documentstyle[preprint,aps]{revtex}
\def\mZ{{\bf Z}}
\def\mR{\relax{\rm I\kern-.18em R}}

\tighten

\begin{document}
\preprint{NSF-ITP-95-154  \qquad gr-qc/95mmnnn}

\title{Criticality and Bifurcation in the Gravitational Collapse
        of a Self-Coupled Scalar Field}
\author{Eric W. Hirschmann\footnote%
        {Electronic address: \tt ehirsch@dolphin.physics.ucsb.edu\hfil}}
\address{Department of Physics\\
        University of California\\
        Santa Barbara, CA 93106-9530}
\author{Douglas M. Eardley\footnote%
        {Electronic address: \tt doug@itp.ucsb.edu\hfil}}
\address{Institute for Theoretical Physics\\
        University of California\\
        Santa Barbara, CA 93106-4030}
\date{\today}

\maketitle

\begin{abstract}

We examine the gravitational collapse of a non-linear sigma model in
spherical symmetry.  There exists a family of continuously self-similar
solutions parameterized by the coupling constant of the theory.
These solutions are calculated together with the critical exponents for black
hole formation of these collapse models.  We also find that the sequence
of solutions exhibits a Hopf-type bifurcation as the continuously
self-similar solutions become unstable to perturbations away from
self-similarity.
\end{abstract}

\narrowtext


\newcommand{\uh}{\hat{u}_{1}}
\newcommand{\fh}{\hat{f}_{1}}
\newcommand{\bh}{\hat{b}_{1}}

\newcommand{\gtwid}
   {\mathrel{\raise.3ex\hbox{$>$\kern-.75em\lower1ex\hbox{$\sim$}}}}
\newcommand{\ltwid}
   {\mathrel{\raise.3ex\hbox{$<$\kern-.75em\lower1ex\hbox{$\sim$}}}}

\section{Introduction}

The last few years have seen a renewed interest in gravitational
collapse, particularly with regard to what numerical relativity is able
to teach us about the general phenomenon.  Choptuik's initial discovery of
criticality and other behavior strikingly similar to that
seen in statistical mechanical systems has suggested a deep
property of the gravitational field equations.

A good deal of recent work has
shown the existence of collapse solutions exactly at
the threshold of the formation of a black hole for a variety of matter
fields.  These include both real and complex scalar fields
\cite{Chop,HE}, vacuum
gravity\cite{AE}, a perfect fluid\cite{EC}, and an axion-dilaton
model from low energy
string theory \cite{EHH}.  In each of these models, some common
behavior emerges.  For example, the growth of the black hole mass just
off threshold is described by a power law relation
\begin{equation}
M_{\rm BH}(p) \propto \cases{
        0,                                              &$p\le p^*$\cr
        (p-p^*)^\gamma,					&$p>p^*$\cr}
\label{bhmass}\end{equation}
where $p$ is any parameter which can be said to
characterize the strength of the initial conditions, and $p^{*}$
is the threshold value, {\it i.e.,} the value for the critical
solution.  The critical exponent $\gamma$ is universal within a
particular class of matter fields.  For example, $\gamma\approx 0.37$
for the real scalar field, $\gamma\approx 0.36$ for perfect fluid
collapse, and $\gamma\approx 0.2641066$ for the axion-dilaton (axiodil)
system \cite{HHS}.
The solutions may also exhibit an echoing behavior in that the
features of the exactly critical solution are repeated on ever
decreasing time and length scales.  This self-similar behavior of the
solutions has been found in both discrete and continuous versions.  In
particular, for vacuum gravity, discrete self-similarity and echoing
are observed, while in fluid collapse, continuous self-similarity
with no echoing emerges.  In scalar field collapse, both types have
been shown to be present.

The main results of this paper unify the discrete {\em vs}
continuous self-similarity known in the above models.  Specifically,
we examine a particular non-linear sigma model which smoothly
interpolates between the complex scalar field model \cite{HE} and
the axion-dilaton
model \cite{EHH} as the value of a certain dimensionless coupling
constant $\kappa$
varies.  We find a family of continuously self-similar solutions
parametrized by $\kappa$.  Using linear perturbation theory,
we study the stability of these solutions, and find
that the sequence of solutions undergoes a bifurcation at
a particular value, $\kappa\approx0.0754$, where the continuously
self-similar solutions go from being stable to being
unstable.   The free complex scalar field ($\kappa=0$) is found to
be on the unstable side of this bifurcation, while the axion-dilaton field
($\kappa=1$) is on the stable side.  This is in agreement with
previous results for both of these matter fields.  Further, we find that
for negative values $\kappa\ltwid-0.28$,  the self-similar
solutions become ever more unstable hinting at the possibility of
further bifurcations, more complicated dynamics, and perhaps even
chaotic behavior in the collapse of these particular models.  Since we
work only in perturbation theory, these latter results are highly
tentative, but they suggest the existence of more exotic behavior
than may have previously been observed.  For this reason, full scale
numerical work on these models would undoubtedly be a very enlightening
undertaking.

Prior to our work, Choptuik and Liebling \cite{Lieb,ChopPriv}
recently studied an apparently different model, namely Brans-Dicke
gravity coupled to a free real scalar field, for various values of
the dimensionless Brans-Dicke coupling constant,
$-3/2<\omega_{BD}<\infty$.   They use a spherical collapse code, and
their main result is a change of stability at $\omega_{BD}\approx0$.
After the continuously self-similar solution was found in the collapse
of an axion-dilaton field\cite{EHH}, they realized that it was
their more general Brans-Dicke model for a particular value of
$\omega_{BD}$.  In fact, we find
that their Brans-Dicke model is
equivalent to some range of our nonlinear sigma model
($\infty>\kappa\ge0$), with $\omega_{BD}=\infty$ corresponding to the
free complex scalar field, and $\omega_{BD}=-11/8$ corresponding to
the axion-dilaton field.  The bifurcation in stability we find here
in linear perturbation theory then coincides with the change of stability
previously found by Choptuik and Liebling;  in particular, we agree
with their result that, for axion-dilaton collapse, the continuously
self-similar critical solution is stable and appears to be the
attractor.  The range $\kappa<0$ is not present in the Brans-Dicke
model, however.

After the research presented here was completed, but before
this paper was posted,
Hamade, Horne, \& Stuart \cite{HHS} reported numerical and
perturbation results on axion-dilaton collapse in spherical symmetry.
Our results in linear perturbation theory agree with theirs
with regard to real modes and critical exponents.  They also find by a
numerical collapse code that the continuously self-similar critical
solution is stable and is the attractor, in agreement with the
work of Choptuik and Liebling;  this is also consistent with our
results below on the complex modes.

The outline of this paper is as follows.  In Section II, we give
general arguments on what kinds of self-coupling of a scalar field
may show new critical phenomena in gravitational collapse;  likely
candidates are the non-linear sigma models.  In Section III, we
introduce the particular non-linear sigma model to be studied
in this paper, and discuss its relationship to matter fields which
have been studied previously.  Section IV introduces the equations of
motion, derives their form in the presence of a continuous
self-similarity, and sketches our numerical approach to solving them.
Section V discusses the perturbation of the continuously
self-similar solutions and the question of stability of these solutions.
Section VI presents our results and conclusions.



\section{Critical Behavior and Self-Interaction}

With the important exception of \cite{AE}, all the work so far on
critical phenomena in gravitational collapse has assumed spherical
symmetry.  In spherical symmetry, there is no gravitational collapse
without matter, from Birkhoff's theorem.  Therefore one might expect
that critical behavior would depend importantly on the model
of the matter.  Indeed, the critical phenomenology and exponents
differ among matter models such as real scalar field, ideal gas,
complex scalar field, axiodil $\ldots$.  However, studying a real
scalar field $\phi$, Choptuik found that inclusion of a nonlinear
interaction term $V(\phi)$ in the action,
\begin{mathletters}\label{scalint}\begin{eqnarray}
 L_{matter}	&=&{1\over2}\nabla^\alpha\phi\nabla_\alpha\phi-V(\phi),\\
 V(\phi)	&\equiv&\mu^2\phi^2/2+\lambda\phi^4/4
\end{eqnarray}\end{mathletters}
made no
difference in the critical solution itself or in its phenomenology.

We can understand this result as follows.  At least in all known
cases, the critical solution is either ``echoing'' (discretely
self-similar) or continuously self similar (CSS -- admitting a homothetic
Killing vector field).  In either case, by dimensional analysis,
the solution cannot depend on any dimensionful parameters.
Here we use dimensional analysis appropriate to classical general
relativity, with a unit of length $\ell$ in some system of units
where Newton's gravitational constant $G\equiv1$.
A scalar field $\phi$ (real or complex) then has dimensions $\ell^0$,
while a Lagrangian must have units $\ell^{-2}$.  It
follows that the parameters $\mu$ and $\lambda$ above have
dimensions different from zero;
in particular, $\mu$ is just the inverse compton wavelength of the particle.
Since these parameters are dimensionful, the critical solution cannot
depend on them, consistent with the numerical results.
\footnote{Choptuik has also tried adding a conformal coupling
$\xi R\phi^2$ to the matter Lagrangian.  In contrast, $\xi$ is dimensionless,
so that that critical solution should depend on it.  This point
deserves more investigation.}

For this reason we turn attention to a different kind of self-coupling,
one which multiplies the kinetic term instead of adding to it.  The
general form is
\begin{equation}
{1\over2}G_{IJ}(\phi^K)\nabla^\alpha\phi^I\nabla_\alpha\phi^J
\end{equation}
where there are now some number $N$ of scalar fields $\phi^I$
($I=1\ldots N$), and where $G_{IJ}$ is some function of the fields,
fixed once and for all to specify the model.  The nonlinear functions
$G_{IJ}$ take the form of a Riemannian metric on the internal
space of the $\phi^I$, the {\sl target space.}  Such models are called
non-linear sigma models (or ``harmonic map" models,
as discussed by Misner\cite{Misner}), and much is known about them in high
energy physics, not least because they often appear in the low energy
limit of superstring theory.  By dimensional analysis,
the scalar fields $\phi^I$ are of dimension $\ell^0$, as is the
target space metric
$G_{IJ}$.  Therefore any parameters appearing in $G_{IJ}$ may
also be taken as dimensionless, and we can expect the critical
solution to depend on them.

What is the simplest nonlinear sigma model we can study?  If
$N=1$ then the matter action can be reduced to that of a
free field by a field redefinition;  a 1-dimensional Riemannian
space is always flat.  So the simplest nontrivial value is $N=2$,
wherein the two real scalar fields can be grouped into a single
complex scalar field $\phi$.  For the target space metric, the
simplest cases are the spaces of constant curvature, namely the
2-sphere, flat 2-space, or the 2-hyperboloid, all with homogeneous
metrics.  This is the model we shall study.


\section{The Model}

We work with a model defined by the following action
\begin{eqnarray}
S = \int d^4 x \sqrt{-g} \left( R -
	{2|\nabla F|^2 \over (1-\kappa|F|^2)^2} \right) .
\label{action}\end{eqnarray}
The complex field $F(x^\mu)$ is a scalar coupled to Einstein gravity
with $\kappa$ a real dimensionless coupling constant,
\begin{eqnarray}
	-\infty<\kappa<\infty.
\label{range}\end{eqnarray}
The model given by Eq.~(\ref{action}) is a
nonlinear sigma model.
As mentioned above, the
target space of the model is a two-dimensional space of constant
curvature.  The curvature of this internal space is proportional to $-\kappa$
so that the space is hyperbolic for $\kappa>0$
and a 2-sphere for $\kappa<0$.  For the particular case $\kappa=1$, our
model
becomes the axion-dilaton (axiodil) field $\tilde{\tau}$
coupled to gravity\footnote{Notation: We use $\tilde{\tau}$ here for the
axiodil field, instead of $\tau$ as we did in \cite{EHH}, to avoid
confusion with  logarithmic time coordinate $\tau$ below.}
\begin{equation}
	F = {1+i\tilde{\tau}\over1-i\tilde{\tau}}.
\end{equation}
It turns out that the value $\kappa=1$ is not affected by
quantum corrections as
it can be protected by extended supersymmetry.
For $\kappa=0$ the model (\ref{action}) reduces
(after a further trivial rescaling
of the field) to the free complex scalar field coupled to gravity.  Thus
this general model smoothly interpolates between the two particular matter
models that we have already considered.  In fact, for
$0<\kappa<\infty$ we find that this nonlinear sigma model is equivalent to
the model of a massless real scalar field coupled to Brans-Dicke theory.
Liebling has recently examined this theory using a version of Choptuik's
adaptive mesh refinement algorithm.  He finds behavior qualitatively similar
to that found by Choptuik for
the real scalar field\cite{Lieb}.  The connection between the
two theories can be seen in the relationship between the Brans-Dicke
coupling constant \cite{Weinberg} $\omega_{BD}$ and our constant $\kappa$
\begin{equation}
	\omega_{BD} = -{3\over2} + {1\over8\kappa},
			\quad 0\le\kappa<\infty.
\label{bd}\end{equation}
This means that the axion-dilaton model ($\kappa=1$) corresponds to
$\omega_{BD}=-11/8$, while the free complex scalar field ($\kappa=0$)
corresponds to $\omega_{BD}=+\infty$.\footnote{As
$\omega_{BD}\rightarrow-3/2^+$,
we have $\kappa\rightarrow+\infty$; however this may be a funny limit.}
For $-\infty<\kappa<0$ the model (\ref{action}) appears not to be
equivalent to any
Brans-Dicke model;  in particular Eq.~(\ref{bd}) does not apply.  The model
behaves in a smooth way as $\kappa$ passes through zero.

Returning to the full model, the field equations in covariant form as
derived from the action
in Eq.~(\ref{action}) are,
\begin{mathletters}\label{covareqns}\begin{eqnarray}
R_{ab} &=& {1 \over (1 - \kappa|F|^2)^2} \left(\nabla_{a}F\nabla_{b}F^{*}
   +   \nabla_{a}F^{*}\nabla_{b}F \right),  \\
\nabla^{a}\nabla_{a} F &=&  {-2\kappa F^{*} \over 1-\kappa|F|^2} \nabla_a F
     \nabla^{a}F .
\end{eqnarray}\end{mathletters}%
In this form, these equations are invariant under a global $U(1)$ group
of transformations for a constant $\Lambda$
\begin{equation}
F' = e^{i\Lambda} F, \quad -\infty < \Lambda < \infty
\end{equation}
and which leave the metric unchanged.

For $\kappa>0$, this model also has an extra global symmetry not
present in general relativity, namely an $SL(2,\mR)$ symmetry that acts on
$F$, but leaves the spacetime metric invariant;  this
is a classical version of the conjectured $SL(2,\mZ)$ symmetry
of heterotic string theory called $S$-duality~\cite{SenRev}.
For the axiodil, $\kappa = 1$, this symmetry acts on $\tilde{\tau}$ as
\begin{equation}
\tilde{\tau} \rightarrow {a\tilde{\tau}+b \over c\tilde{\tau}+d} \; ,
\label{sltwo}\end{equation}
where $(a,b,c,d) \in \mR$ with $ad - bc = 1$, while
leaving $g_{\mu\nu}$ invariant.  The corresponding transformation of
$F$ for general $\kappa>0$ is
\begin{equation}
F \rightarrow {1 \over \sqrt{\kappa}}{\alpha \sqrt{\kappa}F + \beta \over
      \beta^{*} \sqrt{\kappa}F + \alpha^{*}},
\label{sltwof}\end{equation}
where $(\alpha,\beta) \in {\cal{C}}$ with $|\alpha|^2 - |\beta|^2 = 1$.

In the case where $\kappa=0$, the extra global symmetry consists of
translations in the two flat directions of the target space.  Finally,
for $\kappa<0$, the group of motions on the 2-sphere, $SO(3)$,
constitutes the extra global symmetry.

\bigskip

\section{The Continuously Self-Similar Solutions}

We briefly review the process of setting up the equations such that they
are compatible with a continuous
self-similarity.  To begin, we work in spherical
symmetry so the metric can be taken as
\begin{equation}
  ds^2 = (1+u)\left[-b^2dt^2 + dr^2\right] + r^2d\Omega^2
\label{metricrt}\end{equation}
where $b(t,r)$ and $u(t,r)$ are the metric functions.  This is
essentially Choptuik's metric in radial gauge with some minor
redefinitions.
The timelike coordinate $t$ is chosen so that the collapse on the
axis of spherical symmetry happens at $t=0$ and the metric is regular
for $t<0$.

We are interested in finding collapsing solutions of our model.  In
particular we ask whether, as in the complex scalar, axiodil, and fluid
collapse cases, there exist continuously self-similar (CSS)
solutions to these
equations for arbitrary $\kappa$.
That a spacetime admits a continuous self-similarity is described covariantly
by the existence of a homothetic Killing vector field $\xi$ satisfying
\begin{equation}
{\cal L}_{\xi} g_{ab} = \nabla_{a}\xi_{b} + \nabla_{b}\xi_{a} = 2
      g_{ab}  ,
\label{Lxi}\end{equation}
where ${\cal L}$ denotes the Lie derivative.  A coordinate system better
adapted to our assumption of self-similarity
involves the coordinates $z=-r/t$ and $\tau=\ln |-t|$.
In these coordinates, the metric takes the form
\begin{equation}
        ds^2 = e^{2\tau}\left( (1+u)\left[-(b^2-z^2)d\tau^2 +
        2zd\tau dz+dz^2\right] + z^2d\Omega^2 \right),
\label{metric}\end{equation}
and the homothetic Killing vector is then expressed
simply in these coordinates as
\begin{equation}
\xi^{a}\partial_{a} = \partial_{\tau}  .
\label{defxi}\end{equation}

In these coordinates, Eqs.~(\ref{covareqns}) can be written as\footnote{For
completeness, we have included the field equations in the $(t,r)$
coordinates in the appendix.  However, they are not crucial to our
current discussion.}
\begin{mathletters}\label{eqnsinz}\begin{eqnarray}
zu' - \dot{u} & = & {z (u+1) \over \rho^2} \left[ F'(zF'-\dot{F})^{*} +
    F'^{*}(zF'-\dot{F}) \right]  \\
u' &=& {z (u+1)\over \rho^2}\left[|F'|^2 + {1 \over b^2}|zF'-\dot{F}|^2
   \right] - {u(u+1) \over z}  \\
b' & =& {ub \over z}   \\
0 & =& F''\Delta - \ddot{F} + 2z\dot{F}'
     + F'\left[z(u-2) + {b^2 \over z}(u+2) - z{\dot{b} \over b}
    \right] \\
 && \qquad  + \dot{F}({\dot{b} \over b} + 1 - u)
     + {2\kappa \over \rho} F^{*} (\Delta F'^2 + 2zF'\dot{F} - \dot{F}^2 )
\end{eqnarray}\end{mathletters}
where the overdot here means $\partial/\partial\tau$ and the
prime denotes $\partial/\partial z$ and we define the functions
\begin{equation}
\Delta = b^2 - z^2 \quad \rho = 1 - \kappa |F|^2  .
\end{equation}

The boundary conditions we use are that the solution is regular on the
time axis $z=0$ and on the so called similarity horizon
$\Delta=b^2-z^2=0$.  Regularity on the time axis $z=0$ at the center
of spherical symmetry allows us to write the boundary conditions for
the metric functions $b(\tau,z)$ and $u(\tau,z)$ as
\begin{equation}
b(\tau,0) = 1 \quad u(\tau,0) = 0
\label{bcs}\end{equation}
The hypersurface defined by $\Delta=0$ is where the homothetic Killing
vector becomes
null.  As this hypersurface is in the Cauchy development of the
initial
data, we expect everything to be perfectly regular there even though
this is a singular point of Eqs.~(\ref{eqnsinz}).

The existence of the homothetic Killing vector simplifies these
equations somewhat.
For the general collapse problem without self-similarity, the metric
coefficients $u$
and $b$ will be functions of $z$ and $\tau$, but our assumed symmetry
restricts these coefficients to be functions of $z$ alone.  We could also let
the field $F$ be invariant under the action of the vector field $\xi$, but that
would then fail to incorporate the $U(1)$ symmetry which the field equations
also possess.  To allow for more interesting dynamics to occur, we let a
$U(1)$ transformation accompany the scale transformations
(translations in $\tau$).  Infinitesimally, this amounts to
\begin{equation}
{\cal L}_{\xi} F = \xi^{a}\partial_{a}F = i\omega F .
\label{mix}\end{equation}
This allows us to give the form of $F$ under our assumption of
self-similarity as
\begin{equation}
F(\tau,z) = e^{i\omega\tau} f(z) .
\label{field}\end{equation}

The continuously self-similar (CSS) fields are now
\begin{mathletters}\begin{eqnarray}
	F(\tau,z) &=& e^{i\omega\tau} f_{0}(z)  \\
	b(\tau,z) &=& b_0(z)   \\
	u(\tau,z) &=& u_0(z) ,
\end{eqnarray} \end{mathletters}%
where $\omega$ is an eigenvalue, determined by solving the field
equations.  The subscript $_0$ that we have appended denotes
unperturbed values in anticipation of
our eventually perturbing the exactly self-similar solution.

Our equations are now just Eqs.~(\ref{eqnsinz}) with the $\tau$ derivatives of
$u(z)$ and $b(z)$ vanishing, $F$ and $F'$ being replaced by $f_0$ and
$f_0'$, and $\dot{F}$ and $\ddot{F}$ being replaced by $i\omega f_0$
and $-\omega^2 f_0$ respectively.  Note that with $\dot{u}_0=0$, we
can eliminate $u_0'$ and we are left with an algebraic relation for
$u_0 (z)$.
The equations of motion now reduce to\footnote{It is worth pointing
out that our notation here is more
closely
aligned with  what we use in our paper \cite{EHH} on the axiodil, $\kappa=1$,
than the  notation in our papers \cite{HE,HE2} on the complex scalar field,
$\kappa=0$.}
\begin{mathletters}\label{unpert}\begin{eqnarray}
b_{0}' & = & {b_0 u_0 \over z } \\
\Delta_0 f_{0}'' & = & f_{0}'\left(-2i\omega z - z (u_0 - 2) - {b_{0}^2
\over z} (u_0 + 2) - { 4i\kappa\omega z \over \rho_0} |f_{0}|^2 \right) \\
  && \qquad - f_{0} \left(\omega^2 + i\omega (1 - u_0) \right) -
 {2\kappa\over\rho_0} f_0^{*}(\Delta_0 f_0^{\prime{}2} + \omega^2 f_0^2)
\end{eqnarray}\end{mathletters}
where we have defined
\begin{mathletters}\begin{eqnarray}
\Delta_0 & = & b_0^2 - z^2  \\
\rho_0 & = & 1 - \kappa|f_0|^2  \\
u_0 & = & {z^2 \over \rho_0^2} ({1 \over b_0^2}|i\omega f_0 - z
f_0'|^2
    + |f_0'|^2 ) + \\
 && \qquad {z \over \rho_0^2} \left(f_0'(i\omega f_0 - zf_0')^{*} +
   f_0'^{*}(i\omega f_0 - z f_0') \right)
\end{eqnarray}\end{mathletters}
and where the prime now denotes $d/dz$.

The boundary conditions at $z=0$ for the CSS problem now reduce to
\begin{equation}
b_{0}(0) = 1, \quad f_{0} = \mbox{free real constant}, \quad f_{0}'(0) = 0,
\end{equation}
where we have used our $U(1)$ phase symmetry to fix $f_0$ as real.
We define the value of $z$ where $\Delta_0$ vanishes as $z_2$.
As mentioned earlier, we demand regularity  at $\Delta_0(z_2)=0$
and this leads to the additional boundary
conditions
\begin{equation}
b_0(z_2) = z_2 = \mbox{free real const} \quad f_0(z_2) = \mbox{free
complex const}
\end{equation}
with the constant $f_0'(z_2)$ being determined by Eq.~(\ref{unpert}) at
the similarity horizon.

Now with the equations and boundary conditions, we can numerically
integrate these equations.   Once we reduce our
second order ODE to two first order ODEs and include the
real eigenvalue $\omega$ we have five real equations and five real unknowns.
We use our standard technique of solving this
two-point boundary value problem by shooting with an adaptive ODE solver
from both boundary points to a point $z_1$ in the middle.  The free
boundary values
are then found using a Newton's solver for the nonlinear matching
conditions.\cite{NR}

We then follow the CSS solution as $\kappa$ varies, and we find that a CSS
solution exists for
\begin{equation}
	-0.60\ltwid \kappa<+\infty;
\end{equation}
for $\kappa=0,1$ the CSS solution is the same one found in previous work.
Our computations actually only extend to $\kappa\le15$, but the behavior
is smooth and the CSS solutions seem likely to extend all the way
to $\kappa=\infty$.  On the other hand, our calculations of CSS solutions
appear to terminate somehow at $\kappa\approx-0.60$.  We
are unsure what exactly goes wrong there, but we tend to believe that
our numerical routine fails and it is not the case that the CSS
solutions cease to exist for smaller $\kappa$.
It is, however, worth noting that Maison found that his sequence of CSS
gas collapses terminated at a
maximal value $k_{max}
\approx0.88$ where $k$ appears in the equation of state for an Eulerian
fluid $p=k\rho$.
The reason in his case was a change in the nature of the eigenvalues
associated with the singular sonic point.  At $k_{max}$, two of the
eigenvalues degerate.  But we have no evidence that a
similar thing occurs here.

As far as we know, there is only one eigenvalue $\omega$ possible
for the CSS solution for a given $\kappa$;  however we have
not looked very carefully for others.  

We also mention that although we describe the spacetime only up to
the similarity horizon, the spacetime can be continued in these
coordinates to $z=+\infty$.
This corresponds to the spacelike hypersurface $t=0$.  We expect
everything to be regular on this hypersurface
except at the axis of spherical symmetry since it too is in the
Cauchy development of the initial data.
Thus the apparent singularity in our equations at $z=+\infty$
is merely a coordinate singularity.  By changing coordinates, we
can continue the spacetime through $t=0$.
We do not construct this extension here, but as it was possible
to make this continuation for the complex scalar field and the axiodil
cases, we fully expect
that such a construction should be possible\cite{HE,EHH}.

\section{Perturbations and Stability}

As interesting as the CSS solutions are, they do not tell us everything
we would like to know about the gravitational collapse.  After all,
these are the exactly critical solutions $p=p^{*}$ and comprise  a set
of measure zero in the space of initial conditions of the collapse.  To
reach them, the initial conditions must be tuned with exquisite care.
In addition, such things as the critical exponents of the black hole
scaling relation are found only with information gained by collapse
slightly away from the critical solution.

For these reasons, we look to perturbation theory for additional
understanding of the CSS solutions.  It too is not the last word, but it
can shed some light on questions of stability and in particular allow us
to calculate the critical exponents of the black hole growth.

As described in \cite{HE2}, the very construction of a Choptuon involves
stabilization -- a balancing between subcritical dissipation and
supercritical black hole formation with the critical exponent $\gamma$
measuring the strength of this black hole/dissipation instability.
More specifically, for initial data close to, but not exactly on
the critical solution, the
critical solution serves as an intermediate attractor with near-critical
solutions approaching it but eventually running away from it to form
a black hole or dissipate the field to
infinity.
However, in addition to this particular
instability, we would like to know if there are
{\em additional} instabilities which would rather drive the near-critical
solutions completely
away from the Choptuon to another, perhaps very different,
attractor.
Thus by appealing to perturbation
theory, we are looking for both the black hole instability ({\it i.e.} the
critical exponent) and possibly other instabilities indicating
the existence of other, stronger attractors.

So, with the continuously self-similar solutions in hand,
we now carry out a linear perturbation analysis of the CSS solutions,
still in spherical symmetry.
We define the perturbed fields as
\begin{mathletters}\label{pertvars}\begin{eqnarray}
  b(\tau,z)	&\approx& b_0(z) + \epsilon b_{1}(\tau,z) \\
  u(\tau,z)	&\approx& u_0(z) + \epsilon u_{1}(\tau,z) \\
  F(\tau,z)	&\approx& e^{i\omega\tau}\left(f_0(z) + \epsilon
	f_{1}(\tau,z) \right)
\end{eqnarray}\end{mathletters}%
where again, ${}_0$ denotes the 0th order critical solution, ${}_1$
denotes the 1st order perturbation, $\omega$ is
the (unique) eigenvalue of the unperturbed equations (which depends
on the coupling constant $\kappa$), and
where $\epsilon>0$ is an infinitesimal constant, a measure of how
far away the solution is from the critical solution in the space of
initial conditions.  Using Choptuik's terminology, we consider the
supercritical regime for infinitesimal
\begin{equation}
	\epsilon \propto p - p^*.
\end{equation}

We now perturb the Einstein equations through 1st order in
$\epsilon$, to obtain a set of linear partial differential
equations for the perturbed fields $b_1$, $u_1$, $\fh$, in the
independent variables $\tau$, $z$.\@ Following the standard
approach, we Fourier transform the 1st order fields with respect to
the ignorable coordinate $\tau = \log(-t)$,
\begin{mathletters}\label{fourier}\begin{eqnarray}
  \uh(\sigma,z)	&=& \int e^{i\sigma\tau} u_1(\tau,z)d\tau,\\
  \bh(\sigma,z)	&=& \int e^{i\sigma\tau} b_1(\tau,z)d\tau,\\
  \fh(\sigma,z) &=& \int e^{i\sigma\tau} f_1(\tau,z)d\tau;
\end{eqnarray}\end{mathletters}%
throughout, $\hat{}$
will denote such a Fourier transform.  The transform coordinate
$\sigma$ is in general complex.  The 1st order field equations now
become ordinary differential equations (ODE's) in $z$, and under
appropriate boundary conditions, become an eigenvalue problem for
$\sigma$\null.  Solutions of the eigenvalue problem are then normal
modes of the critical solution.  Generally speaking, there will be many
different normal modes $\fh$, each belonging to a different eigenvalue
$\sigma$.  Eigenvalues in the lower half
plane ${\rm Im}\sigma<0$ belong to unstable (growing) normal modes.
Eigenvalues in the upper half $\sigma$ plane
correspond to quasi-normal (dying)
modes of the critical solution.  The eigenvalue $\sigma$ is related
to the critical exponent by $\gamma =
-1/\rm{Im}\sigma$.\cite{KHA,Maison,HE2}

We now want to integrate our equations numerically so we need to
determine
the boundary conditions.  It is important to bear in mind that in
addition to solving the equation for $\fh(\sigma,z)$, we must
also solve
the analogous equation for $\fh(-\sigma^{*},z)^{*}$.  Thus, we will have
two second order ODE's which must be reduced to four first order ODE's,
we will
have a total of six complex equations to integrate.
For the perturbation problem, the boundary conditions at $z=0$ are found
to be
\begin{equation}
\bh(0) = 0, \quad \uh(0) =0, \quad \fh'(\sigma,0) = 0,
\quad \fh'^{*}(-\sigma^{*},0) = 0,
\label{bc01}\end{equation}
\begin{equation}
\fh(\sigma,0) = \mbox{free complex constant}, \quad
\fh^{*}(\sigma^{*},0) = \mbox{free complex constant} .
\label{bc02}\end{equation}
At the similarity horizon, $z=z_2$, the boundary conditions are
as follows.  Both $\bh(z_2)$ and
$\uh(z_2)$ are free complex constants.  Either $\fh(\sigma,z_2)$ or
$\fh'(\sigma,z_2)$ is a free complex constant with the other
describable in
terms of the other boundary conditions at $z_2$.  We chose to let
$\fh'(\sigma,z_2)$ to be free and $\fh(\sigma, z_2)$ fixed as
this facilitated
examining the lower half complex $\sigma$ plane.  The same is true for
the values $\fh'(-\sigma^{*},z_2)^{*}$ and $\fh(-\sigma^{*},z_2)^{*}$.
Counting the eigenvalue $\sigma$, we now have seven pieces of complex
boundary data to go with the six complex equations we need to integrate.
Since the perturbation equations are linear, we expect the solutions to
scale, so the extra piece of data is merely a reflection of the
linearity of the equations.  Solutions will come in families which
will be parameterized by a single complex parameter.
Thus we have an eigenvalue problem which is
well posed and which should yield a discrete spectrum of eigenvalues
$\sigma$.

To solve the 1st order problem we used a Runge-Kutta integrator with
adaptive stepsize as part of a standard two point shooting
method \cite{NR}, shooting from $z=0$ and from both boundaries and
matching in the middle $z=z_{1}$.  For convenience we solved
the 0th order system, Eqs.~(\ref{unpert}), and the 1st order system,
Eqs.~(\ref{pert}) simultaneously with the same steps in $z$.  As
discussed elsewhere, the similarity horizon $z_2$ is a demanding
place to enforce a boundary condition, and a second order Taylor
expansion of the regular solution was used for this purpose.

To solve the 1st order system, we collected all the boundary values
but $\sigma$ into a complex 6-vector
$X\equiv(\fh(\sigma,0),\fh(-\sigma^{*},0)^{*},\bh(z_2),\uh(z_2),
\fh'(\sigma,z_2),\fh'(-\sigma^{*},z_2)^{*})$.
Because the equations are linear, the matching conditions at
$z=z_{1}$ are likewise linear in the boundary values.  A solution
is found when the values at $z_1$ of
$(\bh,\uh,\fh(\sigma),\fh(-\sigma^{*})^{*},\fh'(\sigma),\fh'(-\sigma^{*})^{*})$
upon integrating from $z=0$ match with those found by integrating
from $z=z_2$, for some boundary values $X$.  We can express this
matching condition
\begin{equation}
	A(\sigma) X = 0
\end{equation}
where $A(\sigma)$ is a $6\times6$ complex matrix which is a
nonlinear function of $\sigma$, constructed numerically by
integrations of the 1st order equations, Eqs.~(\ref{pert}), for six
linearly independent choices of boundary values $X$.  The condition
on $\sigma$ for a solution is then
\begin{equation}
	\det A(\sigma)=0.
\end{equation}
Once a value for $\sigma$ was found that satisfies this condition,
the corresponding boundary values $X$ were found as a zero
eigenvector of the matrix $A$; these come in one (complex)
parameter families, as observed above.  Solution of
Eqs.~(\ref{pert}) with boundary values $X$ yields the normal mode
itself.
Now, $A(\sigma)$ has been carefully constructed so that it is a
complex {\it analytic} solution of $\sigma$.  This follows from the
fact that all equations leading to $A$ contain $\sigma$ but not
$\sigma^*$, together with some standard theorems about ODE's.
Moreover, $A(\sigma)$ has no singularities in the closed lower half
$\sigma$ plane.  These properties allow us to use a number of ideas
from scattering theory to study $\det A(\sigma)$.  In particular,
there is a theorem for counting the number $N_C$ of zeros of $\det
A$ within any closed contour $\cal C$ in the closed lower half
$\sigma$ plane:
\begin{equation}
	\Delta_{\cal C}{\rm Arg}\det A = 2\pi N_{\cal C}
\label{count}\end{equation}
where ${\rm Arg}\det A$ is the phase of $\det A$, and $\Delta_{\cal
C}{\rm Arg}\det A$ is the total phase wrap (in radians) around the
closed contour ${\cal C}$, a result similar to Levinson's
theorem for counting resonances in quantum scattering theory.

Furthermore, a conjugacy relation holds,
\begin{equation}
	A^*(-\sigma^*) = A(\sigma),
\end{equation}
which means that $A$ need only be evaluated for ${\rm Re}\sigma\ge0$
in the lower half plane.

The nonlinear equation $\det A(\sigma)=0$ was solved by the secant
variant of Newton's method \cite{NR}.  The equation being
complex-analytic, the 1-complex-dimensional realization of the
method was used, and it performed well.


Since our field equations possess gauge invariance due to general
coordinate invariance and global $U(1)$ phase invariance, some
unphysical pure gauge modes will appear at 1st order, to the extent
that the gauge conditions implicit in our boundary conditions
Eqs.~(\ref{bc01},\ref{bc02}) fail to be unique.

A pure gauge mode arises from an infinitesimal phase
rotation $\phi\rightarrow e^{i\epsilon}\phi$ in the 0th order
critical solution:
\begin{mathletters}\label{gauge1}\begin{eqnarray}
        \bh (z)    &=& 0,\\
        \uh (z)    &=& 0,\\
        \fh (z)       &=& i f_0(z).
\end{eqnarray}\end{mathletters}%
This gives a time independent solution of Eqs.~(\ref{pert}) that
satisfies the boundary conditions; hence it corresponds to an
unphysical mode at $\sigma=0$

Another pure gauge mode results by adding an infinitesimal
constant to time $t\rightarrow t+\epsilon$ at constant $r$ in the
0th order critical solution.  This is possible because our
coordinate conditions, Eqs.~(\ref{bcs}) normalize $t$ to
proper time along the negative time axis $(t<0,z=0)$, but the zero
of time is not specified.  Then the solution is perturbed by
\begin{mathletters}\label{gauge2}\begin{eqnarray}
    b_1(\tau,z) &= {\partial b_0\over\partial t}|_r &= -(z/t)b'(z) =
                                e^{-\tau}zb'(z),\\
    u_1(\tau,z) &= {\partial u_0\over\partial t}|_r &= -(z/t)u'(z) =
                                e^{-\tau}zu'(z),\\
   f_1(\tau,z) &= e^{-i\omega\tau}{\partial(e^{i\omega\tau}f_0)\over
       \partial t}|_r &= e^{-\tau}(-i\omega f_0(z)+z f_0'(z)).
\end{eqnarray}\end{mathletters}%
This pure gauge mode has time dependence $e^{-i\sigma\tau} =
e^{-\tau}$ and so has negative imaginary $\sigma=-i$.

There are also two more gauge modes which appear as a pair on the real
axis.  These come from the addition of an infinitesimal complex constant,
$c$, to our zeroth order solution:
$F\rightarrow F + \epsilon c$.  The perturbed fields are then
\begin{mathletters}\begin{eqnarray}
b_1(\tau,z) & = 0,  \\
u_1(\tau,z) & = 0,  \\
f_1(\tau,z) & = ce^{-i\omega\tau}.  \\
\end{eqnarray}\end{mathletters}%
This mode has a time dependence of $e^{-i\sigma\tau} = e^{-i\omega\tau}$
and so has $\sigma = \omega$.  Of course, since we have
$A^{*}(-\sigma^{*}) = A(\sigma)$, the value $\sigma = -\omega$ will
also solve the equation $\det A=0$
and be the fourth gauge mode.\footnote{When we did a similar
analysis for the complex scalar field, we were insensitive to these
modes since we worked with the derivatives of $\phi(\tau,z)$ so
that these modes vanished
identically.}   Thus there are these four gauge modes
in the $\sigma$ plane and no others.  These modes should appear as
numerical solutions, but are unphysical.

\section{Results and Conclusions}

On integrating and solving for the eigenvalues, $\sigma(\kappa)$, we
found some novel behavior.  We confirmed the existence
of the gauge modes thereby checking the consistency of our method.  We also
found the critical exponent $\gamma(\kappa)$ over the range of $\kappa$
values for which we found a solution.  Figure 1 is a graph of this
exponent as a function of the coupling constant.  As can be seen, the
critical exponent for the CSS solution depends strongly on the value of
the coupling constant.

In addition, we evaluated $\det A(\sigma)$ around a large rectangular
contour in the lower half plane and used Eq.~(\ref{count}) to count the
zeros lying within.  This allowed us to determine if there were
additional modes in the lower half $\sigma$ plane.
Our results were as follows.  We
did find many more modes in the complex $\sigma$ plane.
These additional modes are initially in the upper half plane for large
positive $\kappa$ and approach the real axis as $\kappa$ decreases.
Once one of these modes crosses the axis into the lower half plane we
infer that
the leading normal mode of the CSS solution has a change of stability.
This first occurs at
$\kappa\approx0.0754$.  We thus have the following
\begin{eqnarray}
	0.0754	&\ltwid\kappa<+\infty, \qquad \hbox{CSS stable}\cr
	-0.60	&\ltwid \kappa \ltwid0.0754, \qquad \hbox{CSS unstable}\cr
\end{eqnarray}
This confirms the original discovery by Choptuik and Liebling of a change of
stability at $\omega_{BD}\approx0$; from Eq.~(\ref{bd}) the value would be
$\omega_{BD}\approx0.158$\null.\cite{Lieb,CL}  Note that these
results are in good agreement with earlier work.  The CSS solution
for the complex
scalar field ($\kappa = 0$) was shown to be unstable by a similar
analysis \cite{HE2}
while the CSS solution for the axion-dilaton field ($\kappa=1$)
was recently shown in
\cite{HHS}
to be the attractor in gravitational collapse and hence agrees with what we
have found here, namely that the solution found in
\cite{EHH} is stable.  An important question is if the CSS solution
becomes unstable at $\kappa\approx 0.0754$, what is the attractor for the
collapse.  Our conjecture, borne out by collapse calculations of
Choptuik and Liebling, is that the attractor between $0<\kappa\ltwid 0.0754$
({\it i.e.} $\omega_{BD} \gtwid 0$) is the more dynamically interesting
discretely self-similar (DSS)
or echoing solution analogous to the
echoing solution originally seen by Choptuik in the collapse of a real
scalar field.

Since everything in our model is smooth at $\kappa=0$, as we
decrease $\kappa$ below zero,
we expect the relevant attractor for the collapse
to continue to be the echoing solution.
However, the above mentioned unstable mode turns out not to be the
only mode to move into the lower half plane.
We have evidence for more modes going unstable by $\kappa\approx-0.28$.
We have been able to construct these perturbation modes and they appear
to be legitimate solutions of the perturbation
equations and not numerical artifacts.
The presence of these additional modes suggests that the model becomes
ever more unstable, that is more nonlinear, as $\kappa$ decreases.
So what happens in gravitational collapse as $\kappa$ decreases below
$\approx-0.28$?  The CSS solution will certainly
not be the attractor and the existence of additional unstable
modes may trigger further
bifurcations in the echoing solution.
Since our calculations are limited to perturbation theory, we can not
say this with certainty, but our expectation is that the echoing
solution will become unstable and bifurcate into an even more
dynamically complicated solution.  One way to determine what happens
here with greater assurance
would be to take a numerical solution for the DSS solution
and perform a
perturbation analysis.  That this would be feasible is suggested by
Gundlach's results in which he calculates the echoing solution as an
eigenvalue problem resulting from the assumption of discrete
self-similarity in the collapse of a real scalar field.\cite{Gund}
However, a more direct approach would be
to perform a full scale numerical collapse
calculation in order to understand what is going on in this regime.

In this paper, we have combined a few of the previously studied models
of gravitational collapse into a single model of a self-coupled complex
scalar field.  The model is paratmeterized by a single coupling constant
$\kappa$.  In Table 1, we give a summary of some of the key values of
$\kappa$.  As the value of the coupling constant decreases, the
continuously self-similar solution which we find undergoes a change in
stability.  For the regime where the CSS solution is unstable, we
believe that the attractor for gravitational collapse is an echoing
and discretely self-similar solution.  This change in stability which
occus near $\kappa=0.0754$ looks like a ``Hopf Bifurcation", as it is
known in the dynamics literature\cite{suz}.
As $\kappa$ continues to decrease, we find evidence for additional
instabilities in the model, suggesting that there exists at least
another bifurcation of the collapsing solution.  From the lore on other
dynamical systems, this further conjectural bifurcation might lead to a
doubly periodic attractor, or might lead to full blown dynamical chaos
in gravitational collapse.  Additional work will be able to
determine whether that is indeed the case.

\bigskip
\indent\vtop{\advance\hsize by -1in\parindent=0pt
{\bf Table 1.\quad} Range of the model, its relation to the Brans-Dicke
model, critical exponents, and stability.

\halign{
$#$\hfil&\qquad$#$\hfil&\quad$#$\hfil&\quad#\hfil\cr
\multispan4{\hrulefill}\cr
\hbox{Nonlinear} &\hbox{BD/scalar}\cr
\hbox{sigma, }\kappa	&\omega_{BD}	&\gamma		&Stability of CSS\cr
\multispan4{\hrulefill}\cr
+\infty		&-3/2		&\ltwid0.14(?)	&Stable?\cr
10.0		&-1.4875	&0.1469		&Stable\cr
1		&-11/8		&0.2641		&Stable\cr
\ltwid 0.0754   &\gtwid0.158    &\gtwid0.373	&Becomes Unstable\cr
0		&+\infty	&0.3871		&Unstable\cr
\ltwid -0.28	&n/a		&\gtwid0.435	&Becomes Mucho Unstable?\cr
\ltwid-0.60	&n/a		&		&(Don't know if CSS exists)\cr
\multispan4{\hrulefill}\cr
}\hss}
\bigskip

\acknowledgements

This research was supported in part by the National Science
Foundation under Grant Nos.~PHY89-04035 and PHY90-08502, and parts
were carried out at the Aspen Center for Physics.  We are grateful
to Matt Choptuik and Jim Horne for enlightening communications.

\appendix
\section{ }

In this appendix, we simply list some of the equations that seemed too
cumbersome to burden the main portion of the paper with.  We have the
general collapse equations for our model (\ref{action}) in the usual
$(t,r)$ coordinates.

\begin{mathletters}\begin{eqnarray}
\dot{u} & = & {r(u+1) \over \rho^2} \left[\dot{F}^{*}F' + \dot{F}F'^{*}
   \right]                            \\
u' & = & {r(u+1) \over \rho^2} \left[|F'|^2 + {1 \over b^2}|\dot{F}|^2
    \right]    - {u(u+1) \over r} \\
b' & = & {ub \over z}  \\
0 & = & r^2 ({1 \over b}\ddot{F} - {\dot{b} \over b^2} \dot{F}) -
   (r^2 bF'' + 2rbF' + r^2 b'F') \\
  && \qquad - {2\kappa r^2 \over \rho} F^{*}( bF'^2 -{1 \over
                b} \dot{F}^2 )
\end{eqnarray}\end{mathletters}%
where $\dot{ }$ means $\partial/\partial t$ and $^{\prime}$ means
$\partial/\partial r$, and where
$$
\rho = 1 - \kappa |F|^2.
$$

The Eqs.~(\ref{eqnsinz}) when perturbed as given in Eqs.~(\ref{pertvars})
become our
original set as well as the following Fourier-transformed first order
equations
\begin{mathletters}\label{pert}\begin{eqnarray}
\lefteqn{z\uh' + i\sigma\uh  = } \\
 & & {z(u_0 + 1) \over \rho_0^2}
     \biggl[  f_0'\left(z\fh' - i(\omega + \sigma^{*})\fh\right)^{*} +
           f_0'^{*}\left(z\fh' - i(\omega - \sigma)\fh\right)  \\
  & & \qquad\qquad\qquad + \fh'(zf_0' - i\omega f_0)^{*} +
              \fh'^{*}(zf_0' - i\omega f_0)   \biggr]   \\
  & &  + {z(u_0+1) \over \rho_0^2}
       \left({\uh \over u_0 +1} - {2 \hat{\rho}_1
                                    \over \rho_0}\right)
     \left[  f_0'(zf_0' - i\omega f_0)^{*} +
             f_0'^{*}(zf_0' - i\omega f_0)    \right]  \\
\uh' & = & {z(u_0+1) \over \rho_0^2}
   \biggl[  f_0'\fh'^{*} + f_0'^{*}\fh'
          -{2\bh\over b_0^3} |zf_0'-i\omega f_0|^2   \\
  & & \qquad       + {1\over b_0^2}
   \left(  (zf_0' - i\omega f_0)(z\fh' - i(\omega+\sigma^{*})\fh)^{*}
      + (zf_0' - i\omega f_0)^{*}(z\fh' - i(\omega-\sigma)\fh)  \right)
                   \biggr]  \\
  & &  + {z(u_0 +1) \over \rho_0^2}
      \left({\uh \over u_0 +1} - {2 \hat{\rho}_1
                                     \over \rho_0} \right)
   \left( |f_0'|^2 + {1\over b_0^2}|zf_0' - i\omega f_0|^2 \right) \\
\bh' &=& {1 \over z }(u_0 \bh + u_1 b_0)  \\
0 & = & \fh''\Delta_0
    + \fh'\left\{ 2iz\left[(\omega-\sigma)
         + {2\kappa\omega\over\rho_0} |f_0|^2\right]
         + {1 \over z} \left[ z^2 (u_0-2) + b_0^2 (u_0+2) \right]
         + {4\kappa\Delta_0\over\rho_0} f_0^{*} f_0' \right\} \\
& & + \fh\biggl\{ (\omega-\sigma)
         \left[ (\omega-\sigma+i(1-u_0))
        + {4\kappa\over\rho_0} f_0^{*} (izf_0' + \omega f_0) \right]\\
   & & \qquad\qquad  +{2\kappa^2\over\rho_0^2} f_0^{*{}2}
  \left[ b_0^2 f_0^{\prime{}2} +  (iz f_0' + \omega f_0)^2 \right] \biggr\} \\
& & + \fh^{*} {2\kappa\over\rho_0^2} \left\{ b_0^2 f_0^{\prime{}2} +
          (iz f_0' + \omega f_0)^2 \right\}    \\
& & + \bh\left\{2b_0 f_0'' + {4\kappa b_0\over\rho_0} f_0^{*} f_0^{\prime{}2} +
     {\sigma \over b_0} (\omega f_0 + izf_0') + {2 b_0
      \over z}(u_0+2) f_0'\right\} \\
& & + \uh \left\{ -i\omega f_0 + z f_0' + {b_0^2\over z} f_0'
      \right\}
\end{eqnarray}\end{mathletters}


\begin{figure}
\caption{This is a graph of the critical exponent, $\gamma$, of the
continuously self-similar solution as a function of $\kappa$.
}
\label{fig1}
\end{figure}

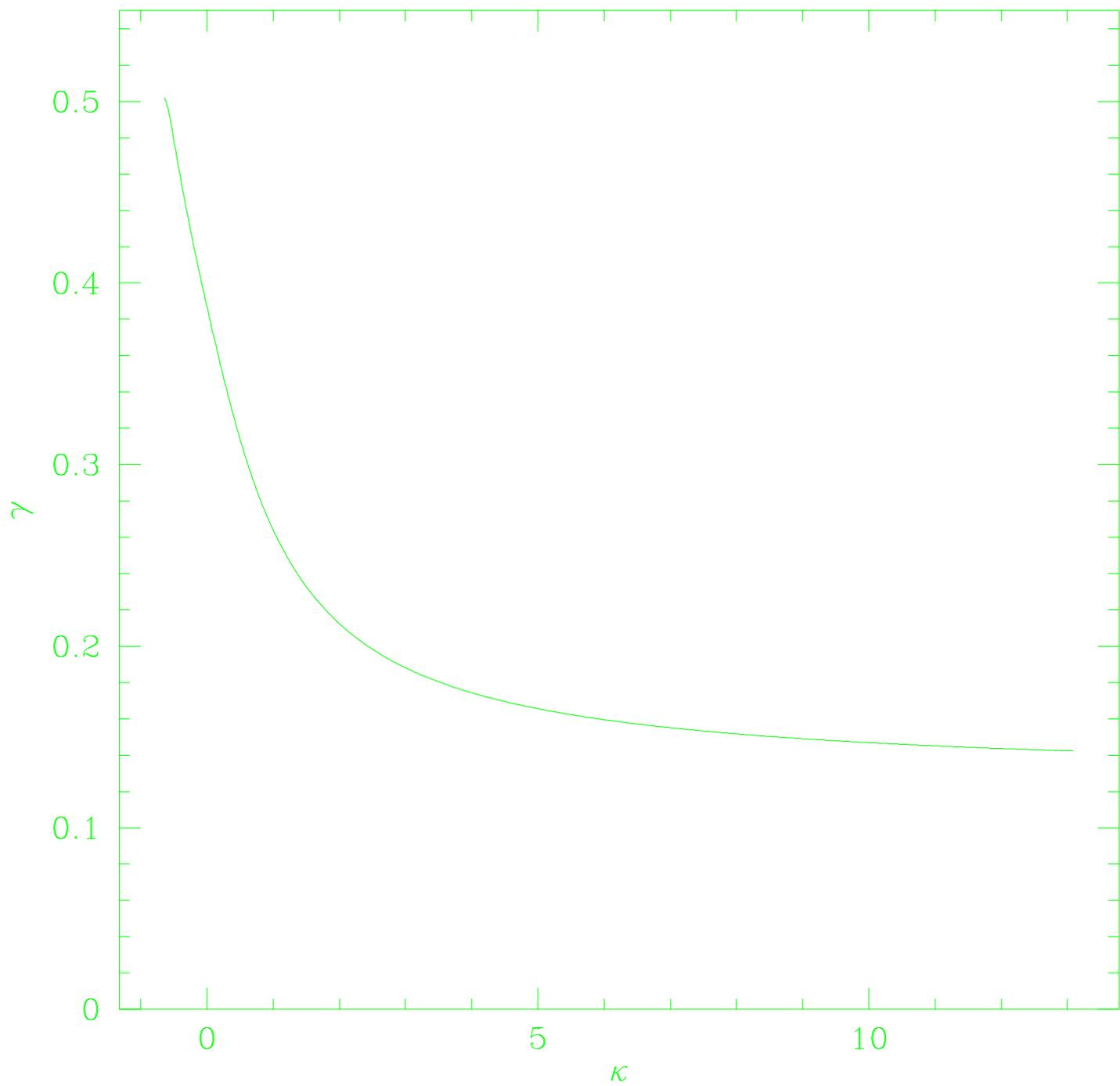
\end{document}